\newcommand{\vecp}{\mbox{$\mathbf{p}$}}

\newcommand{\vecv}{\mbox{$\mathbf{v}$}}
\newcommand{\vece}{\mbox{$\mathbf{e}$}}

\newcommand{\vecE}{\mbox{$\mathbf{E}$}}
\newcommand{\vecB}{\mbox{$\mathbf{B}$}}
\newcommand{\rhoD}{\rho_{\mbox{\tiny D}}}
\newcommand{\vD}{v_{\mbox{\tiny D}}}
\documentclass[12pt,a4paper]{article}
\usepackage[applemac]{inputenc}

\begin{document}

\title{Catalyzing Fusion with Relativistic Electrons}
\author{Hanno Ess\'en\\
Mechanics, KTH}
\date{January 1997}

\maketitle

\begin{abstract}
The idea here is to use large relative velocities of electrons and
nuclei in accelerator beams to increase the probability of fusion.
The function of the electrons is to both screen the positive
charge and to produce an increased parallel pinching current. The
increase in reaction probability is estimated using the Darwin
magnetic interaction energy approach.
\end{abstract}

In order to get fusion of, say, deuterons, one normally requires
very high temperatures. The reason for this is twofold: firstly a
large kinetic energy is needed for the particles to penetrate the
Coulomb barriers, and secondly high speeds are needed to get
appreciable reaction rates in spite of very small cross sections.
The well known problem with this is that the high temperature
makes high density and confinement difficult to achieve. Another
problem is that as the speeds go up the cross sections go down.
The purpose here is to present a radically different idea of
achieving fusion, an idea that starts from the observation that at
relativistic speeds the Coulomb repulsion can be balanced by a
magnetic attraction.

The following well known calculation illustrates the basic facts.
Consider a beam of charged particles moving along a straight line
with speed $v$ and constant charge density $\rho_0$ within some fixed
radius of the line and zero outside. From $\nabla\cdot\vecE =4\pi
\rho_0$ we then get
\begin{equation}
E_r = 2\pi \rho_0 r
\end{equation}
for the radial and only component of the electric field. From
$\nabla\times\vecB = 4\pi \rho_0 \vecv/c$ we similarly get
\begin{equation}
B_{\varphi} = 2\pi \rho_0 \frac{v}{c}r
\end{equation}
for the azimuthal and only component of the magnetic field. Together
these now give for the Lorentz force on a particle of charge $e$ in
the beam
\begin{equation}
\label{eq.lor.force.charge.beam}
F_r = e\left(E_r - \frac{v}{c} B_{\varphi}\right) = 2\pi
\left(1-\frac{v^2}{c^2}\right) e\rho_0 r.
\end{equation}
We see that the Coulomb self repulsion of the beam goes to zero as
$v\rightarrow c$.

The above result in itself is, of course, not useful for fusion
purposes. Firstly, the cost of accelerating two deuterons to
relativistic speeds is far greater that the energy from a fusion
reaction between them. Secondly, by transforming back to the rest
frame of the particles in the beam, we see that they are not likely
to fuse more often just because they have high speed relative to an
irrelevant observer. Both these problems are solved, and this is the
crucial idea here, if a beam of relativistic electrons is injected
into a moderate energy beam of deuterons. To maximize the relative
(and thus `real') speed the electrons should be injected with a
velocity opposite to that of the deuterons.

Let us go back to the calculation above for this situation. If we
manage to make the particle densities equal, $\rhoD
=|\rho_{\rm e}|$, we get $\rho_0 =\rhoD + \rho_{\rm e}=0$
and thus zero electric field: $E_r=0$. For the magnetic field, on the
other hand, we get
\begin{equation}
B_{\varphi}=2\pi\rhoD \left(\frac{ \vD+ |v_{\rm e}| }{c}
\right) r
\end{equation}
since the two currents contributing are assumed to be in the same
direction. The Lorentz force on a deuteron is now
\begin{equation}
F_{r\mbox{\tiny D}} =- e  \frac{\vD}{c} B_{\varphi} =- 2\pi
\left(\frac{ \vD^2+\vD|v_{\rm e}| }{c^2}\right)
e\rhoD r,
\end{equation}
and that on an electron is
\begin{equation}
F_{r\rm e} = e  \frac{v_{\rm e}}{c} B_{\varphi} =- 2\pi
\left(\frac{ |v_{\rm e}| \vD+ v^2_{\rm e} }{c^2}\right)
e\rhoD r.
\end{equation}
The Lorentz force is now seen to strive to contract the beam. Since we
assume that $\vD \ll |v_{\rm e}|$ the force on the electrons is
much greater than on the deuterons. As soon as the electrons have
contracted to smaller radius than the deuterons, however, there will be
a Coulomb force, from them, acting to contract the deuteron beam.

Due to the attraction of parallel currents, alias the `pinch' effect,
the beam of electrons with velocity opposite to that of the positive
ions, leads to an automatic confining action on the combined beam.
So far we have looked at the beam as smeared out charge
densities. We must now consider the forces between the individual
particles in the beam.

There is no known relativistic expression that can be used so we will
consider the Darwin Hamiltonian which is known to be correct to order
$(v/c)^2$. According to the Darwin Hamiltonian  the
interaction energy of two charged particles, $i$ and $j$, is
\begin{equation}
\label{eq.darwin.interaction}
{\cal V}_{ij} =  \frac{q_i q_j}{r_{ij}}-\frac{q_i q_j[\vecp_i \cdot
\vecp_j + (\vecp_i\cdot \vece_{ij})(\vecp_j\cdot \vece_{ij})]}{2c^2
m_i m_j r_{ij}}.
\end{equation}
Here $\vece_{ij}$ is the unit vector from particle $i$ to $j$.
It is assumed that one can replace $\vecp_i/m$ by $\vecv_i$. This
gives:
\begin{equation}
{\cal V}_{ij} =  \frac{q_i q_j}{r_{ij}}\left(1-\frac{\vecv_i \cdot
\vecv_j + (\vecv_i\cdot \vece_{ij})(\vecv_j\cdot
\vece_{ij})}{2c^2}\right).
\end{equation}
We again see that the Coulomb interaction is reduced if the particles
are moving in the same direction, just as in our first study of the
charged beam. Now, however, if we consider the deuterons in the rest
frame of the electrons, which should be relevant since the deuterons
are moving among the electrons, we find a real effect. For
two deuterons moving, along the same line, with speed $v$ relative to
the electrons, we get
\begin{equation}
\label{red.coulomb.rep}
{\cal V}_{ij} =  \frac{e^2}{r_{ij}}\left(1-\frac{v^2}{c^2}\right).
\end{equation}
This indicates that for sufficiently rapid electrons we get rid of the
Coulomb repulsion.

It is clearly not correct to use the Darwin interaction energy, which
is a relativistic correction to the classical interaction, when
the speeds become highly relativistic. There is reason, however, to
believe that the indicated qualitative effect, a reduction of the
Coulomb repulsion, is a real one and persists also in the fully
relativistic case. It is also interesting to note the similarity
of equation (\ref{red.coulomb.rep}) with equation
(\ref{eq.lor.force.charge.beam}), in which no assumption about small
speed was made.

Let us now take equation (\ref{red.coulomb.rep}) seriously. We are
thus considering the deuterons as moving at relativistic speeds
relative to the frame defined by the electrons of the beam. Let us
consider two of these deuterons. Even if they both have relativistic
speeds, their relative speed is not particularly great. It is their
relative speed that determines the effect of their interaction. They
will thus repel each other with a reduced Coulomb force according to
formula (\ref{red.coulomb.rep}). They also attract each other via the
strong force and this can be described qualitatively by a
Yukawa potential. The total interaction potential energy of two
deuterons a distance $r$ apart can thus, now be taken to be
\begin{equation}
\label{red.coulomb.rep.plus.yukawa}
{\cal V}(r) = -A\frac{\exp(-\lambda r)}{r} +
\left(1-\frac{v^2}{c^2}\right)\frac{e^2}{r}.
\end{equation}
Here $A$ and $\lambda$ are constants. In what follows we will use
atomic units ($\hbar=e=m_{\rm e}=1$). Reasonable values to be used in
estimates are then $A=137$ and $\lambda=10000$, and we will use these
below.

We will thus assume that the deuterons interact via the potential
energy (atomic units)
\begin{equation}
\label{red.v.dep.pot}
{\cal V}_v (r) = -137\frac{\exp(- r/10^{-4})}{r}+\frac{\theta(v)}{r}.
\end{equation}
where
\begin{equation}
\theta(v)\equiv 1-\frac{v^2}{c^2}.
\end{equation}
 The question now is what
values of $v$ are reasonable to use. The energy gain from fusion
reactions between deuterons are tabulated in table \ref{table.energies}.
The combined energy of the deuterons and the electrons may not exceed
what can be gained from the fusion reactions if we are to make an
energy profit.

\begin{table}

\begin{tabular}{|l|r|} \hline
Reaction & Energy yield \\
\hline D + D $\rightarrow\; ^3\! {\rm He + n}$ &  3.27  MeV  \\
\hline D + D $\rightarrow\; {\rm T + p}$  &  4.03  MeV  \\
\hline  D + D $\rightarrow\; ^4\! {\rm He}+\gamma$ &  23.85  MeV  \\
\hline  D + p $\rightarrow\; ^3\! {\rm He}+\gamma$ &  5.49  MeV  \\
\hline D + T $\rightarrow\; ^4\! {\rm He + n}$ & 17.59  MeV  \\
\hline
\end{tabular}

\caption[energies]{\small
\label{table.energies}
This table lists the energy yields of some relevant fusion
reactions.}
\end{table}

We see that the available energy is strongly dependent on which of
the three first  reactions of table \ref{table.energies} that occurs.
In ordinary collision reactions between deuterons the two first
reactions dominate completely. The reason is that the high relative
speed of the deuterons does not give time for an electromagnetic
process to occur. The two first reactions involve the strong force
and this force has much smaller time scales. In this application,
however, we do not need high relative velocity since we instead lower
the Coulomb barrier. There is then strong reason to expect that the
third reaction will be much more common. Even if the first are more
common there is then hope that secondary (the fourth and fifth of
table \ref{table.energies}) reactions will occur with the released
protons and tritium ions. In any case one should not have to put in more
than, say, 20 Mev, of energy into the electrons. Note carefully that
the electrons are not consumed by the process; they can be circulated
and reused unless they are scattered or otherwise lost from the beam.

\begin{table}

\begin{tabular}{|c|c|c|c|c|c|c|c|c|} \hline
$E_{\rm e}$ & $v/c$ & $\theta(v)$ & $r_{\rm max}$ &
${\cal V}_v(r_{\rm max})$  & $E$  & $r_-(E)$  & $r_+(E)$  & $P(E,v)$
\\ \hline
 0  &  0  & 1 &  .00070  & 1250.10 &0.01 & .00049 & 100.0 & $10^{-1166}$
\\ \cline{6-9}
 MeV  &    &  &    &  &  0.1 & .00049 & 10.00 & $10^{-366}$
\\  \hline
4.60&.995 &0.01 &.00121 &7.635 &0.01 &.00095 & 1.00 & $6\cdot 10^{-12}$
\\  \cline{6-9}
MeV   &    &  &    &  &0.1 &.00095 & 0.100 & $6\cdot 10^{-4}$
\\ \hline
20.& .9997 & .0006 & .00151 & 0.372 &0.01 & .00123 & .0600 &  $0.27$
\\  \cline{6-9}
 MeV  &    &  &    &  &0.1 & .00126 & .0060 & $0.80$   \\
\hline
\end{tabular}

\caption[param]{\small \label{table.param} Parameters referring to
the potential of formula (\protect\ref{red.v.dep.pot}) for three
different kinetic energies ($E_{\rm e}$) of the catalyzing
electrons. The fraction of the speed of light of the electrons and
the Coulomb repulsion reduction parameter $\theta=1-v^2/c^2$ is
also listed. $r_{\rm max}$ is the position of the maximum of the
potential and the value of the potential at this maximum is given
in the following column. $E$ is the relative kinetic energy of the
deuterons (atomic units), $r_-$ and $r_+$ the corresponding
classical turning points. Finally $P$ is a semiclassical estimate
of the tunnelling probability i.e.\ of the reaction probability.}
\end{table}

We now estimate the probability that the deuterons will tunnel
through the Coulomb barrier of the potential
(\protect\ref{red.v.dep.pot}). The probability will depend on
$\theta(v)=1-v^2/c^2$ and on the relative kinetic energy $E$ of
the deuterons. A semiclassical estimate of the tunnelling
probability is given by
\begin{equation}
P = \exp\left(-\frac{2}{\hbar} \int_{r_-}^{r_+} |p| {\rm d}r \right)
\end{equation}
where
\begin{equation}
p(r) = \sqrt{2m_{\mbox{\tiny D}} [ E-{\cal V}_v (r)] }.
\end{equation}
Six different values of $P$ for $\theta=1, 0.01, 0.0006$, and for
$E=0.1, 0.01$ (atomic units) for each $\theta$, are given in table
\ref{table.param}. It is clear from the table that the tunnelling
probability goes from essentially zero to physically interesting
values when $\theta$ takes values corresponding to relativistic
electrons.

Are these tunnelling probabilities realistic and relevant? If they
are it should be easy to achieve fusion with the method outlined
here. Relevant questions are: what density of Deuterium ions and
electrons are technically possible in the initial beams? How much
does this density increase due to the self contraction when the
beams meet? How large a fraction of the Deuterium ions that have
not fused can be recirculated? How large a fraction of the
relativistic electrons can be recirculated? The answer to these
questions will determine the feasibility, economy, and future
usefulness of the method for controlled fusion proposed above.

\end{document}